\documentstyle[prl,aps,psfig]{revtex}
\newcommand{\be}{\begin{equation}}
\newcommand{\ee}{\end{equation}}
\newcommand{\bea}{\begin{eqnarray}}
\newcommand{\eea}{\end{eqnarray}}

\def\tq{${\bf q}$}

\begin{document}
\title{Exact Calculation of the Spatio-temporal Correlations in the
Takayasu model and in the \tq-model of Force Fluctuations in Bead Packs}
\author{R. Rajesh$^1$ and Satya N. Majumdar$^{1,2}$}
\address{
{\small 1. Department of Theoretical Physics, Tata Institute of Fundamental
Research, Homi Bhabha Road, Mumbai 400005, India}\\
{\small 2. Laboratoire de
Physique Quantique (UMR C5626 du CNRS), Universit\'e Paul Sabatier,
31062 Toulouse Cedex, France} }
\date{\today}
\maketitle
\widetext
\begin{abstract}
We calculate exactly the two point mass-mass correlations in arbitrary
spatial dimensions in the aggregation model of Takayasu. In this model,
masses diffuse on a lattice, coalesce upon contact and adsorb unit mass
from outside at a constant rate. Our exact calculation of the variance of
mass at a given site proves explicitly, without making any assumption of
scaling, that the upper critical dimension of the model is $2$. We
also extend our method to calculate the spatio-temporal correlations in a
generalized class of models with aggregation, fragmentation and injection
which include, in particular, the $q$-model of force fluctuations in bead
packs. We present explicit expressions for the spatio-temporal force-force
correlation function in the $q$-model. These can be used to test the
applicability of the $q$-model in experiments.

\end{abstract}

\section{Introduction} 

It is well established that interacting many body non-equilibrium systems,
evolving via the dynamics of their microscopic degrees of freedom, can
reach a large variety of steady states in the limit of large time. Among
these steady states, those characterized by power law distributions of
different physical quantities, over a wide range of parameter space, have
received a lot of attention in the last decade. The phenomenon of the
emergence of power law distributed steady states without fine tuning of
external parameters has been dubbed self organized criticality (SOC)
\cite{btw}.  The concept of SOC has been very useful in understanding
scale invariance in a large number of physical systems including sand
piles, driven interfaces, river networks and earthquakes.  One of the early 
models of SOC is the mass aggregation model proposed by Takayasu\cite{tak}. 
This simple lattice model describes a system in which masses diffuse, coalesce
upon contact and adsorb unit mass from outside at a constant rate. In the
limit of large time, the mass distribution evolves into a time-independent
form with a power law tail for large mass, in all dimensions. This
time-independent mass distribution was computed exactly in one dimension
and within mean field theory\cite{tak}.  The appearance of a power law
distribution without fine tuning of parameters and the fact that the power
law exponent can be computed exactly makes the Takayasu model (TM) one of
the simplest analytically tractable models of SOC.

TM also has close connections \cite{dhar1} to other models of statistical
mechanics such as the Scheidegger river network model\cite{scheidegger},
the directed abelian sandpile model\cite{dhar2} and the voter
model\cite{liggett}. TM is exactly equivalent to Scheidegger's river model
which had been proposed to explain theoretically the observed drainage
patterns of river catchment area, in particular Hack's law which
relates the catchment area to the length of the river\cite{hack}.  Though
more sophisticated models have been proposed to explain Hack's law,
Scheidegger's model is a simple stochastic model which explains the law
with fair accuracy.  In the case of the directed abelian sandpile model
the probability of having an avalanche of size $m$ turns out to be
identical to the mass distribution $P(m)$ in the TM. The voter model space
time trajectory is very similar to that of the TM except that the
processes proceed in the opposite direction in time.  Thus, a complete
solution of the TM would also help in understanding  these related models 
better.

There has been a recent revival of interest in the TM as simple modifications
of the model led to the understanding of a variety of systems including force
fluctuations in granular media such as bead packs\cite{coppersmith},
various aspects of river networks\cite{rinaldo},
particle systems in one dimension with long range hops\cite{krug} and
generalized mass transport models\cite{RM}. In addition, two other natural
generalizations of the TM were found to display nontrivial nonequilibrium
phase transitions in the steady state\cite{MKB1,MKB2}.

Even though the single site mass distribution function in the TM has been
computed exactly in $1$-dimension and in the mean field limit, the spatial
and temporal correlations between masses have remained an open question.
In fact, Takayasu and Takayasu, in their recent review article
\cite{tak1}, have commented on the difficulty of computing the spatial
mass-mass correlation function both analytically and numerically.  In this
paper, we calculate exactly both the spatial and temporal mass-mass
correlation functions in the TM in all dimensions. The calculation of the
variance of mass in all dimensions also settles rigorously that the upper
critical dimension of the TM, beyond which the mean field exponents are
correct, is $2$.

Our technique can also be used to calculate the spatio-temporal two-point
correlations in a class of models which are generalizations of the TM.  
In particular we calculate exactly the two-point force-force correlation
function in the $q$-model of force fluctuations in bead
packs\cite{coppersmith,liu}. If experiments can be devised to measure
these spatio-temporal correlations in bead packs, then our exact results 
would be useful for comparison.

The paper is organized as follows. In section II, we define the TM and
briefly review the earlier results on this model. We also review the known
results for the $q$-model of force fluctuations in bead packs and then
summarize our main results.  In section III, we study the
spatio-temporal correlations in the TM in one dimension. In subsection
III-A, we compute exactly the equal-time mass-mass correlations between
two points in space for all times. We also perform a Monte Carlo
simulation to compute this correlation function numerically and find
excellent agreement between numerical and exact results. In subsection
III-B, we solve exactly the temporal autocorrelation and derive its large
time behavior. In section IV, we study the spatio-temporal mass-mass
correlations in arbitrary dimensions.  In section V, we study exactly the
spatio-temporal correlations in the generalized $q$-model. We present
the explicit results for the correlation functions in the
$q$-model when the distribution of transmitted fractions of weights are
uniform. We also provide numerical evidence for the equal-time correlations.
Finally, we conclude in section VI with a summary and outlook.  

\section{Model and Results}
\subsection{Takayasu Model}

For simplicity we define the TM on a one dimensional lattice with periodic
boundary conditions. Generalization to higher dimensions is
straightforward. Each site of the lattice has a nonnegative mass variable.
Given a certain configuration of masses at time $t$, the system evolves
via the following dynamics. Evolution at each discrete time step consists
of two moves : (1) with probability $1/2$ each mass hops to its right and
with probability $1/2$ it stays at the original site and (2) a unit mass
is added to each site.  The first move corresponds to the diffusion of
masses while the second move corresponds to injection of unit masses from
outside.  If the diffusion move (1) results in two masses coming to the
same site, then the total mass at the site simply adds up. Thus, the 
evolution of the masses is described by the stochastic equation,
\be
m_i(t+1)=m_i(t)(1-r_i)+m_{i-1}(t) r_{i-1}+1,
\label{two}
\ee
where the $r_i$'s are independent and identically distributed random
variables taking values $0$ or $1$ with equal probability $1/2$, and 
$m_i$ is the mass at site $i$. While the
first move (1) tends to create big masses via diffusion and aggregation,
the second move (2) replenishes the lower end of the mass spectrum. The
competition between the two, leads in the long time limit, to a
time-independent single-site mass distribution with a power law tail for
large mass.  This happens irrespective of the initial condition. For
convenience, one can start from an initial configuration that has zero
mass at each site.

Note that the dynamics of the TM defined above is parallel, i.e., all
sites are updated simultaneously in every time step. Alternately one could
define the model in continuous time where the mass at every site hops to
the right with rate $p$ and injection of unit masses at every site occurs
with rate $1$. It turns out that the large distance and long time
behaviors of the TM are insensitive to the particular type of dynamics
used. This is in contrast to other recently studied generalized mass
models\cite{krug,RM} where the steady state mass distribution depends non
universally on the type of dynamics. It turns out that while the parallel
version of the TM is convenient for establishing the mapping to other
models, the continuous time version is sometimes easier for the purpose of
calculations. We will therefore use either of the two versions whichever
is more convenient in a particular situation. For example, in one
dimension we will calculate the mass-mass correlation function with
parallel dynamics while for arbitrary dimensions $d$, we will use the
continuous time version as significant simplifications occur in that case.

We now briefly review the earlier results on the TM. Takayasu and
coworkers\cite{tak,tak1} originally showed that the probability $P(m,t)$
that a site has mass $m$ at time $t$ approaches a time-independent form in
the long time limit $t\to \infty$. This time-independent distribution was
shown to have a power law tail, $\sim m^{-\tau}$ for large $m$ and the
exponent $\tau$ was computed exactly\cite{tak} in $d=1$ ($\tau_{1d}=4/3$)
and within mean field theory ($\tau_{mf}=3/2$).  It was also shown that
for large $m$ and large but finite $t$, the distribution satisfies a
scaling form\cite{tak2,MS} 
\be 
P(m,t)\sim \frac{1}{m^{\tau}} f(\frac{m}{t^{\delta}}), 
\label{one} 
\ee 
where the exponent $\delta$ is related to $\tau$ via the simple scaling 
relation, $\delta=1/(2-\tau)$\cite{MS}. Recently Swift et. al.\cite{maritan} 
have argued that in $d$ spatial dimensions, $\tau=2(d+1)/(d+2)$ for $d<2$ 
and $\tau=3/2$ for $d>2$.  In $d=2$, they argued that in the limit $t\to
\infty$, $P(m)\sim m^{-3/2}(\log m)^{1/2}$ for large $m$ indicating that
$2$ is the upper critical dimension of the TM.  It is also known\cite{tak}
that the power law distribution of mass is stable with respect to
fluctuations in the initial conditions and is insensitive to whether the
particles hop symmetrically in space or not. 

While the single site mass distribution in the TM is known analytically as
mentioned above, the spatio-temporal correlations between masses are far
from being understood. Recently Takayasu and Takayasu have pointed out
that while there exist spatial correlations in TM, they are difficult to
compute not only analytically but even numerically\cite{tak1}.  In
addition there is no rigorous derivation of the upper critical dimension
$d_c$ of the model. While equivalence with the voter model would predict
$d_c=2$\cite{dhar2}, early numerical simulations\cite{tak} suggested
$d_c=4$. The argument of Swift et. al.\cite{maritan} that $d_c=2$ is also
not rigorous as it relied on a scaling ansatz for the age distribution of
particles which looked plausible but was not proved.

\subsection{The ${\bf q}$-Model of Force Fluctuations}

The $q$-model was proposed\cite{liu,coppersmith} as a simple scalar model
to understand the distribution of forces observed in real three
dimensional bead packs\cite{liu}. This model assumed that the force chains
observed in experiments were due to inhomogeneity in packing leading to
unequal distribution of weights supported by a bead. Ignoring the spatial
correlations between inhomogeneities, the model considered a regular
lattice of sites, each containing a bead of mass unity. The total weight
on a given bead at a layer is transmitted randomly to $2$ nearby beads in
the layer underneath (In the original version of the model, $N$ adjacent
beads were considered). Let $m(i,t)$ be the weight supported by a bead at
site $i$ at depth $t$ ($t$ is the layer index). Then the transmission of
weights can be represented via the stochastic equation, Eq. (\ref{two}) 
where it is assumed, for convenience, that the successive layers in the
$t$ direction are shifted by one lattice unit to the right. The injection
term $1$ represents the weight of a bead itself (assuming that all beads
have the same weight unity) and $r_i$ represents the fraction of the
weight that is transmitted from a given bead to its descendent in the next
layer to the right. The only difference with the TM is that, in the $q$
model, $r_i$'s are independent and identically distributed random
variables drawn from a uniform distribution over $[0,1]$. Indeed one can
study a general stochastic equation such as Eq. (\ref{two}) where $r_i$'s
are independent and identically distributed random variables drawn from a
general distribution $f(r)$ over $[0,1]$\cite{coppersmith}. TM is a
special case with $f(r)={1\over 2}\delta_{r,0} +{1\over 2}\delta_{r,1}$.
Similarly $q$-model with uniform distribution is another special case when
$f(r)=1$.

For distributions of the form, $f(r)=r^n(1-r)^n/B(n+1,n+1)$ (where $n$ is
a positive integer and $B(m,n)$ is the Beta function), Coppersmith et. al.
showed\cite{coppersmith} that the joint probability distribution of the
normalized weight variables, $v_i=m_i/t$, factors in the limit $t\to
\infty$, i.e., $P(v_1,v_2,v_3\ldots)=\prod_i P(v_i)$ as $t\to \infty$. The
uniform distribution falls in this category as it corresponds to $n=0$.
Thus the correlations between normalized weights vanishes in the $t\to
\infty$ limit and the single point weight distribution was shown to be,
$P(v)=a(n) v^{n+1}\exp (-2nv)$ for all $v$ where $a(n)$ is independent of
$v$\cite{coppersmith}. Experiments on bead packs measured the force
distribution on the bottom layer of the pack ($t\to \infty$) and the
results were found to be in agreement with the $q$-model results with
$n=0$, i.e., with uniform distribution, $f(r)=1$\cite{liu}.

While the spatial correlations between the normalized weights vanish in 
the $t\to \infty$
limit, they are expected to be nonzero at finite depth $t$. We will show
later that Claudin et. al.\cite{Claudin} stated incorrectly that for the
uniform distribution, the correlation is zero at any finite depth.  
Claudin et. al. also calculated\cite{Claudin} the equal-depth correlation
function in a continuum version of the $q$-model for generic distribution
of the fractions and found a rather structure less correlation function.
This is not surprising because they made the assumption
that the beads are massless. Our exact calculation for the discrete
$q$-model in this paper shows that for nonzero bead mass, the equal-depth
correlation function has a very interesting scaling behavior characterized
by a universal scaling function which is independent of initial conditions
for short-ranged initial conditions. Besides, we also compute exactly the
nontrivial temporal correlations between masses in the vertical direction.
These temporal correlations have not been computed for the $q$-model
before.

\subsection{ New Results}
The new results that we obtain in this paper can be summarized as follows: 

\noindent 
(1) For the TM, we calculate exactly the equal-time mass-mass
correlation function $C(r,t)=\langle m_0(t)m_r(t)\rangle -\langle
m_0(t)\rangle \langle m_r(t)\rangle$ between two spatial points separated
by a distance $r$ in all dimensions. We show that in the scaling limit
$r\to \infty$, $t\to \infty$ but keeping $r/{\sqrt t}$ fixed, $C(r,t)=
-t^{\gamma} G(r/{\sqrt t})$ where $\gamma=2$ for $d<2$ and
$\gamma=(3-{d\over 2})$ for $d>2$ and the scaling function depends on $d$.
In $d=2$, there are additional
logarithmic corrections. In $d=1$, we also compute the scaling function
$G(y)$ explicitly.

\noindent 
(2) Putting $r=0$ limit in the explicit expression for $C(r,t)$
allows us to compute the on-site variance, $\langle m^2(t)\rangle$
exactly. For large $t$, we find $\langle m^2\rangle\sim t^{(4+d)/2}$ in
$d<2$, $\langle m^2\rangle \sim t^3/{\log t}$ in $d=2$, and $\langle
m^2\rangle \sim t^3$ in $d\geq 3$. For $d=1$, this result was already
derived by Takayasu et. al.\cite{tak} but for $d\neq 1$ it is a new
result. This therefore proves rigorously, without any assumption of
scaling, that the upper critical dimension (beyond which mean field
exponents are correct) of the TM is $d_c=2$.

\noindent
(3) We also study exactly the normalized unequal time correlation
function, $A_r(t,\tau)=\langle X_0(t) X_r(t+\tau)\rangle/{\sqrt {{\langle
X_0^2(t)\rangle}{\langle X_r^2(t+\tau)\rangle}} }$ where
$X_r(t)=m_r(t)-\langle m_r(t)\rangle$, in all dimensions in the TM.  The
normalized autocorrelation function is obtained by putting $r=\tau/2$ in
$A_r(t,\tau)$. This is because each mass in the TM has a net drift
velocity equal to $1/2$ to the right.  We show that the autocorrelation
function, $A_{\tau/2}(t,\tau)\sim {\tau}^{-d/2}h(\tau/t)$ in the scaling
limit, $\tau \to \infty$, $t\to \infty$ but keeping $\tau/t$ finite. In
$d=1$, we derive the scaling function $h(y)$ explicitly.

\noindent
(4) We also calculate exactly the correlations between forces at two
different points (both equal depth and unequal depth) in the
$q$-model\cite{coppersmith} of force fluctuations in bead packs for
arbitrary distributions of the fractions of weights. These
correlations have so far not been measured experimentally. 
But if experiments can be performed in future,
then our exact results will be useful for validation of the $q$-model.

\section{Correlations in One Dimension}
In this section, we calculate exactly the spatio-temporal correlation
function in the TM in one dimension. Even though the evolution equation
for the single point probability distribution of mass $P(m,t)$ involves
the joint two point probability distribution function $P(m_1,m_2,t)$, the
evolution equation for the two point correlation involves only other two
point correlation functions. This simplifying aspect makes the correlation
function analytically tractable in the TM.

The parallel dynamics of the TM in $1$-d is represented by the stochastic
equation (\ref{two}), namely, $m_i(t+1)=m_i(t)(1-r_i)+m_{i-1}(t)
r_{i-1}+1$. If $r_i=1$ at time $t$, the mass $m_i$ at site $i$ jumps to
its right neighbour while $r_i=0$ indicates that it stays at site $i$.  
The hopping of the mass at site $i$ to $i+1$ is described by the first
term while the second term accounts for the mass at $i-1$ hopping onto
site $i$.  The last term $1$ indicates the injection of unit mass from
outside at every time $t$. Averaging Eq. (\ref{two}) over all possible
histories (starting from a zero mass initial configuration) we immediately
get $\langle m\rangle (t)=t$.

\subsection{Equal Time Correlations in One Dimension}
The evolution equation for the equal-time correlation function between two
space points $i$ and $j$ can be written down by multiplying Eq. (\ref{two})
by $m_j(t+1)$ and then taking an average over all possible histories. Due
to the translational invariance in an infinite lattice, this correlation
function depends only on the difference $|i-j|$. Denoting $x=i-j$ and
using the translational invariance, we find the connected part of the
correlation function at time $t$, $C_x(t)=\langle m_0(t)m_x(t)\rangle
-t^2$, obeys the equation,
\be
C_x(t+1)=\frac{1}{4}( C_{x+1}+2 C_x+C_{x-1})+\frac{1}{4} (C_0+t^2) (2
\delta_{x,0}-\delta_{x,1}-\delta_{x,-1}).
\label{three}
\ee
In obtaining the above equation, we have used the fact that $r_i$ and
$m_j$ at time $t$ are independent of each other for all $i$ and $j$. The
Eq. (\ref{three}) can be solved exactly for arbitrary initial condition by
the generating function method. It turns out, however, that the solution
at large time $t$ becomes asymptotically independent of the initial
condition as long as the initial condition is short ranged. Without any
loss of generality, we therefore start from the simplest initial condition
when the mass is $0$ at every site. Let $F(q,u)=\sum_{x=1}^{\infty}
\sum_{t=0}^{\infty}C_x(t)q^x u^t$. Multiplying Eq. (\ref{three}) by $q^x
u^t$ and summing over $x$ and $t$, one can express $F(q,u)$ in terms of
${\tilde C}_1(u)=\sum_{t=0}^{\infty}C_1(t)u^t$ and ${\tilde
C}_0(u)=\sum_{t=0}^{\infty}C_0(t)u^t$. ${\tilde C}_1(u)$ can further be
expressed in terms of ${\tilde C}_0(u)$ from Eq. (\ref{three}) by putting
$x=0$, multiplying by $u^t$ and summing over $t$. Thus, finally we get the
following expression for $F(q,u)$,
\be
F(q,u)=\frac{q\left[ u^2 (1+u) (1-q)-2 (1-u)^4
{\tilde C}_0(u)
\right] }{(1-u)^3 \left[ 4 q -u (1+q)^2 \right]},
\label{four}
\ee
where ${\tilde C}_0(u)$ is yet to be
determined. We determine ${\tilde C}_0(u)$ by noting that $F(q,u)$ has two
poles $q_{1,2}=(2-u \pm 2 \sqrt{1-u})/u$. For positive values of $u$,
$|q_1|>1$ while $|q_2|<1$. This would imply that for fixed time, $C_x(t)$
will blow up exponentially as $|q_2|^x$ for large $x$. Since this can not
happen, the numerator on the right hand side of Eq. (\ref{four}) must also
vanish at $q=q_2$ in order to cancel the pole. This immediately determines
${\tilde C}_0(u)$,
\be
{\tilde C}_0(u)=\frac{u (1+u)
(1-\sqrt{1-u})}{(1-u)^{7/2}}.
\label{five}
\ee
Substituting ${\tilde C}_0(u)$ in Eq. (\ref{four}), the generating
function $F(q,u)$ is then fully determined, 
\be
F(q,u)=\frac{-u (1+u)}{(1-u)^3}\sum_{n=1}^{\infty}\left(
\frac{u q}{(1+\sqrt{1-u})^2}\right)^n.
\label{six}
\ee
Let us denote ${\tilde C}_x(u)=\sum_{t=0}^{\infty}C_x(t){u}^t$.
The coefficient of $q^x$ for $x\geq 1$ can be easily pulled out from
Eq. (\ref{six}) to yield
\be
{\tilde C}_x(u)=\frac{-u
(1+u)}{(1-u)^3}\frac{u^x}{(1+\sqrt{1-u})^{2 x}}.
\label{seven}
\ee
Note that it is evident from the above expression that $C_x(t)=0$ for
$x\geq t$.

In order to derive the explicit expressions for $C_x(t)$ for $x\geq 0$,
we need to invert the discrete Laplace transforms ${\tilde C}_x(u)$.
We first derive $C_0(t)$ explicitly by computing
the coefficient of ${u}^t$ in the expression of ${\tilde
C}_0(u)$ in Eq. (\ref{five}), 
\be
C_0(t)=\frac{2 t (2 t+1)(4 t+1)}{4^t 15}\frac{(2 t)!}{(t!)^2} -t^2.
\label{eight}
\ee
This therefore gives us an exact expression for the on-site mass variance
$C_0(t)$ for all $t$. Taking the large $t$ limit
in Eq. (\ref{eight}) we get,
\be
C_0(t)=\frac{16 t^{5/2}}{15\sqrt{\pi}}+O(t^2).
\label{nine}
\ee
We note that since Eq. (\ref{one}) implies that $\langle m^2\rangle \sim
t^{(3-\tau)/(2-\tau)}$, we
get $\tau=4/3$ in 1-dimension. This therefore constitutes an alternate
method to derive $\tau=4/3$ in $1$ dimension.

In order to derive $C_x(t)$ for $x>0$, we need to calculate the
coefficient of ${u}^t$ on the right hand side of Eq. (\ref{seven}).
For arbitrary $x$, this is somewhat hard. However it is easy to derive the
asymptotic behavior of $C_x(t)$ for large $x$ and large $t$ but keeping
$x/{\sqrt t}$ fixed. By taking $u \to 1$ limit in Eq. (\ref{seven})
and after a few steps of algebra, we find that in this scaling limit
\be
C_x(t)=-t^2 G(x/\sqrt {t}), 
\label{ten}
\ee
where the scaling function $G(u)$ is universal, i.e., independent of the
initial condition as long as the initial condition is short ranged and is
given by
\be
G(y)=32 \int_{y}^{\infty}d x_1 \int_{x_1}^{\infty}d x_2
\int_{x_2}^{\infty}d x_3
\int_{x_3}^{\infty}d x_4\, {\rm erfc}(x_4).
\label{eleven}
\ee
The complementary error function is defined as, ${\rm {erfc}}(y)={2\over
{\sqrt \pi}}\int_y^{\infty}{\exp}(-x^2)dx$. The above integrals can be done
to derive an explicit expression for the scaling function,
\be
G(y)=\frac{1}{3} \left[ (3+12 y^2 +4 y^4) {\rm {erfc}}(y)-{2\over {\sqrt \pi}}
y (5+2 y^2) e^{-y^2}\right]. 
\label{twelve}
\ee
We also performed a numerical
simulation of the TM on a one dimensional lattice with periodic boundary
condition. In Fig. 1, we show the scaling plot of the connected
part of the correlation function. The data at different times, when
scaled as in Eq. (\ref{ten}) collapse onto a single scaling function
which is in excellent agreement with the analytical result given by 
Eq. (\ref{twelve}).

\subsection{Temporal Correlations in One Dimension}
In this subsection we compute exactly the two time correlations in the TM
in one dimension.  Let us first define, $X_x(t)=m_x(t)-\langle
m_x(t)\rangle =m_x(t)-t$. We then define the general two time correlation
function as $D_x(t,\tau)=\langle X_0(t)X_x(t+\tau)\rangle$. It is also
useful to define the normalized two time correlation function,
\be
A_x(t, \tau)={ {\langle X_0(t)X_x(t+\tau)\rangle}\over 
{ \sqrt { {\langle X_0^2(t)\rangle}{\langle X_x^2(t+\tau)\rangle} }} }.
\label{norm1}
\ee
Clearly then, $A_x(t,\tau)=D_x(t,\tau)/{\sqrt {C_0(t)C_0(t+\tau)}}$ where
$C_0(t)$ is just the on-site mass variance whose exact expression is given
by Eq. (\ref{eight}) of the previous subsection and we have also used the
translational invariance of the equal-time correlation function. Thus we
just need to evaluate $D_x(t,\tau)$ which can be done exactly as follows.

From Eq. (\ref{two}), it is easy to show that
the function $D_x(t,\tau)$ evolves as a function of $\tau$ for fixed $t$ as,
\be
D_x(t,\tau +1)=\frac{1}{2} \left(D_x (t,\tau)+D_{x-1}(t,\tau)\right) ,
\label{thirteen}
\ee
starting from the initial condition $D_x(t,0)=C_x(t)$, where $C_x(t)$ is
the equal-time correlation function computed already in the previous
subsection. Let $H(k,t,\tau)=\sum_{x=-\infty}^{\infty} D_x(t,\tau) e^{i k
x}$. Note that here we used the $x$ summation from $-\infty$ to $\infty$
as opposed to $0$ to $\infty$. This is because $D_x(t,\tau)$ is not equal
to $D_{-x}(t,\tau)$ for $\tau>0$. They become equal only for $\tau=0$ due
to translational invariance. From Eq. (\ref{thirteen}) we get
\be
H(k,t,\tau)=H(k,t,0)\left(\frac{1+e^{ik}}{2}\right)^{\tau}.
\label{fourteen}
\ee
Note that $H(k,t,0)=\sum_{x=-\infty}^{\infty} C_x(t)e^{ikx}$ and
$C_{-x}(t)=C_{x}(t)$ as translational invariance holds for equal-time
correlation function.

By inverting the Fourier transform in Eq. (\ref{fourteen}), we get a
simple expression for $D_x(t,\tau)$ in terms of the equal-time correlation
functions, 
\bea 
D_x(t,\tau)&=&\frac{1}{2 \pi} \int_0^{2 \pi} dk H(k,t,\tau) e^{-i k x}
\nonumber\\ 
&=& \frac{1}{2^{\tau}}\sum_{m=0}^{\tau}{\tau \choose m} C_{x-m}(t). 
\label{fifteen} 
\eea

In order to calculate the auto-correlation function, we note that in the
TM, the masses have a net drift velocity, $v=1/2$ towards the right. This
is because of the definition of the model: in one time step, a mass either
stays at its own site with probability $1/2$, or hops to the neighbour on
the right with probability $1/2$. Thus, to calculate the proper
auto-correlation function, one has to compute it in the moving frame which
is shifting towards right with uniform velocity $1/2$. Hence the correct
auto-correlation function would be given by $D_{\tau/2}(t,\tau)$.  
Putting $x=\tau/2$ in Eq. (\ref{fifteen}) and taking the transform,
${\tilde D}_{\tau/2}(u,\tau)=\sum_{t=0}^{\infty} D_{\tau/2}(t,\tau)u^t$,
we get
\be
{\tilde D}_{\tau/2}(u,\tau)=\frac{1}{2^{\tau}}\left[ {\tau \choose
\tau/2}{\tilde C}_0(u)+2 \sum_{m=\tau/2+1}^{\tau}{\tau \choose
m}{\tilde C}_{m-\tau/2}(u)\right],
\label{sixteen}
\ee
where we have used the symmetry ${\tilde C}_x(u)={\tilde C}_{-x}(u)$.
Using the exact expressions for ${\tilde C}_x(u)$ from Eqs. (\ref{five})
and (\ref{seven}) of the previous subsection in Eq. (\ref{sixteen}) and
after some steps of algebra, we get
\be
{\tilde D}_{\tau/2}(u,\tau)=\frac{1}{2^{\tau}} {\tau \choose
\tau/2} \frac{u (1+u)}{(1-u)^{7/2}}-\frac{u(1+u)}{(1-u)^3
u^{\tau/2}}+\frac{u(1+u)}{(1-u)^{5/2}u^{\tau/2}} \sum_{i=0}^{\tau/2-1} {2 i
\choose i} \left( \frac{u}{4}\right) ^{i},
\label{seventeen}
\ee
where we have used the combinatorial identity \cite{gradshteyn},
\be
\sum_{j=n/2+1}^{n}{n \choose j} k^j=\frac{(1+k)^n}{2}-\frac{k^{n/2}}{2} {n
\choose n/2}-\frac{(1-k) (1-k)^{n-1}}{2}\sum_{i=0}^{n/2-1}{2i \choose i}
\left(\frac{k}{(1+k)^2}\right)^i.
\ee

In order to analyze Eq. (\ref{seventeen}), we first put $u=1-s$ and note
that the equation allows a scaling limit when $s\to 0$ and $\tau\to
\infty$ but the product $s\tau$ remains fixed.  In terms of time
variables, this scaling limit corresponds to $\tau\to \infty$, $t\to
\infty$ but keeping the ratio $\tau/t$ fixed. In this limit, we find
\be
{\tilde D}_{\tau/2}(s,\tau)={1\over {s^3}}g_1(s\tau),
\label{scale1}
\ee
where the scaling function is given by
\be
g_1(y)= {\sqrt {8\over {\pi}}}\left[ {1\over {\sqrt y}}-{\sqrt {\pi\over 2}}
e^{y/2}{\rm erfc}(\sqrt {y\over 2})\right].
\label{scale2}
\ee
In terms of time variables $\tau$ and $t$ in the scaling limit, $\tau\to
\infty$ and $t\to \infty$ but keeping $\tau/t$ fixed, we get the following
expression by inverting the Laplace transform in Eq. (\ref{scale1}),
\be
D_{\tau/2}(t,\tau)=t^2\left[ { {16{\sqrt 2}}\over {15\pi}}
{\sqrt {t\over \tau}}-h_1({\tau\over t})\right],
\label{nineteen}
\ee
where
\be
h_1(y)=\frac{(1+y/2)^2}{3 \pi}\left( 6 \sin^{-1}\left(\frac{1}{\sqrt{1+y/2}}
\right)-\frac{\sqrt{y}(10+3 y)}{\sqrt{2}(1+y/2)^2} \right).
\label{twenty}
\ee

Using the above result and the exact large $t$ behavior of $C_0(t)$ from
Eq. (\ref{nine}) in Eq. (\ref{norm1}), we finally obtain the scaling
behavior of the normalized autocorrelation function in the scaling limit
mentioned above,
\be
A_{\tau/2}(t,\tau)={1\over {\sqrt \tau}}h\left ( {\tau\over {t}}\right ),
\label{scale3}
\ee
where the scaling function is given by,
\be
h(y)= {\sqrt {2\over \pi}} {1\over {(1+y)^{5/4}}} \left [ 1- { {5{\sqrt {2y}}}
\over {32}} (1+y/2)^2\left( 6\sin^{-1}\left(\frac{1}{\sqrt{1+y/2}}
\right)-\frac{\sqrt{y}(10+3 y)}{\sqrt{2}(1+y/2)^2} \right)\right ].
\label{scale4} 
\ee
The function $h(y)\to {\sqrt {2\over \pi}}$ as $y\to 0$ and $h(y)\approx
{\sqrt {8\over {49\pi}}}y^{-9/4}$ as $y\to \infty$. Thus, for large $y$,
the scaling function decays as a power law with an exponent $9/4$.

We remark that the above scaling behavior holds only in the limit when
$\tau\to \infty$, $t\to \infty$ but the ratio $\tau/t$ is kept fixed. In
other limits, it is also possible to investigate the detailed behavior of
the autocorrelation function by analyzing the exact equation
(\ref{seventeen}).

\section{Correlations in Arbitrary Dimensions}
In this section we study the two-point spatio-temporal correlations in the
TM in an arbitrary spatial dimension $d$. As mentioned in section II, it
turns out that for general $d$, equations for the correlations simplify
considerably for the continuous time version of the TM. In this version,
every mass hops with rate $p$ to each of its $d$ nearest neighbors in the
positive direction, and aggregates with the mass present at the hopped
site. In addition, injection of unit mass occurs at every lattice site
with rate $1$.

The evolution of the mass $m(x_1,x_2\ldots x_d,t)$ in a small time  
interval $\Delta t$ can be represented by the equation,
\be
m(\{x_i\},t+\Delta t)=\sum_{j=1}^d {r_j}^{-}m(x_1,\ldots,x_j-1,\ldots,x_d,t)+
(1-\sum_{j=1}^d {r_j}^{+})m(\{x_i\},t) +I(\{x_i\},t),
\label{dconti}
\ee
where ${r_i}^{\pm}$'s are independent and identically distributed variables, 
with distribution $f(r)=p\Delta t\delta_{r,1}+(1-p\Delta
t)\delta_{r,0}$ and indicate the hopping events of the particles. 
The random variable $I(\{x_i\},t)$ denotes the event
of injection and is drawn from the distribution, $P(I)= \Delta t \delta_{I,1}
+(1-\Delta t)\delta_{I,0}$ independent of
the ${r_i}^{\pm}$'s.  

\subsection{Equal-Time Correlations}

From the equation (\ref{dconti}) of evolution of the masses, one can
easily write down the evolution equation for the two-point correlation
function in continuous time. Multiplying Eq. (\ref{dconti}) at two
different space points and neglecting terms of order $O({\Delta t}^2)$
and higher, we find that the two-point correlation function,
$C(\{x\},t)=\langle m(0,\ldots,0,t)m(x_1,\ldots,x_d,t)\rangle-t^2$,
evolves as
\bea
\frac{d}{dt}C(\{x\})&=&-2p d C(\{x\}) +
p \sum_{j=1}^{d}\sum_{m=\pm 1}C(x_1,\ldots,x_j+m,\ldots,x_d)
+\delta_{x_1,0}\ldots \delta_{x_d,0}\nonumber \\
&+&p\left( C(\{0\})+t^2\right) (2 d \delta_{x_1,0}\ldots\delta_{x_d,0}
-\sum_{j=1}^{d}\delta_{x_1,0}\ldots\delta_{x_j,\pm
1}\ldots\delta_{x_d,0}),
\label{twentyone}
\eea
where we have suppressed the $t$ dependence of $C(\{x\},t)$ for
notational convenience and also used translational invariance of 
$C(\{x\},t)$.  The Fourier transform $G(\{k\},t)= \sum_{\{x\}=-\infty}
^{\infty} C(\{x\})\exp(i\sum_{j=1}^{d} k_j x_j)$ 
then evolves as,
\be
\frac{d}{dt}G(\{ k \} )=2 p\left(G(\{k\})-C(\{0\})-t^2\right) 
\left(-d +\sum_{j=i}^{d} \cos (k_j) \right)+1.
\label{twentytwo}
\ee
Taking Laplace transform with respect to $t$, we get
\be
F(\{k\},s)=\frac{s^2+2 p\left(s^3 g_0(s)+2\right)\left(
d-\sum_{j=1}^{d}\cos(k_j) \right)}
{s^3\left[s+ 2 p\left( d-\sum_{j=1}^{d}\cos(k_j)\right)\right]},
\label{twentythree}
\ee
where $F(\{k\},s)=\int_0^{\infty}G(\{k\},t)e^{-s t}dt$ and
$g_0(s)=\int_0^{\infty}C(0,\ldots,0,t)e^{-s t}dt$ which is yet to be 
determined. 
We determine $g_0(s)$
by noting that $g_0(s)=\frac{1}{(2\pi)^d} \int_0^{2 \pi} \ldots
\int_0^{2\pi}F(\{k\},s)dk_1\ldots d k_d$. Integrating Eq. (\ref{twentythree})
with respect to the $k_i$'s we get the following expression for
$g_0(s)$,
\be
g_0(s)=\frac{1}{s^4}(s^2-2 s+\frac{2}{I(s)}),
\label{twentyfour}
\ee
where 
\be
I(s)=\frac{1}{(2 \pi)^d}\int_{0}^{2 \pi}dk_1 \ldots \int_{0}^{2\pi}d k_d
\frac{1}{s+2 p\left(d-\sum_{j=1}^{d}\cos(k_j)\right)}.
\label{twentyfive}
\ee

The small $s$ behavior of $I(s)$ can be easily evaluated by analyzing
the integral in Eq. (\ref{twentyfive}). We find that as $s\to 0$,
\begin{eqnarray}
I(s) &\sim & s^{-(1-{d\over 2})} \,\,\,\,\, {\rm for}\,\,\,\, d<2  \nonumber \\
     &\sim & -\log (s) \, \,\,\,\, {\rm for}\,\,\, d=2 \nonumber \\
     & \sim & {\rm constant} \,\,\,\,\, \, {\rm for}\,\,\, d>2. 
\label{Is}
\end{eqnarray} 
Substituting $I(s)$ in the expression for $g_0(s)$ in Eq. (\ref{twentyfour}) 
and inverting the Laplace transform, we find that the
on-site variance for large $t$ behaves exactly as,
\begin{eqnarray}
C(0,\ldots,0,t) &\sim & t^{{{4+d}\over 2}} \,\,\,\,\, {\rm for}\,\,\,\, d<2  
\nonumber \\
     &\sim & {t^3\over {\log (t)}} \, \,\,\,\, {\rm for}\,\,\, d=2 \nonumber \\
     & \sim & {t^3} \,\,\,\,\, \, {\rm for}\,\,\, d>2.
\label{var}
\end{eqnarray} 
Note that the results in Eq. (\ref{var}) are exact results for large $t$ and 
does not assume any scaling behavior.
This result clearly proves {\em rigorously} that the upper critical dimension 
of TM is $d_c=2$. 

In addition, if we assume that the on-site mass distribution scales as, 
$P(m,t)\sim m^{-\tau}f(mt^{-\delta})$ for large $m$ and large $t$, we
get the following results for $\tau$ and $\delta$. The first moment, 
$\langle m\rangle\sim t$ gives $\delta=1/(2-\tau)$\cite{MS}.
The second moment scales as, $\langle m^2\rangle \sim t^{(3-\tau)\delta}$. 
Using the exact results for variance from Eq.
(\ref{var}),
we get, $\tau= 2(d+1)/(d+2)$ for $d<2$, $\tau=3/2$ for $d>2$ and $\tau=3/2$ 
in $d=2$ with additional logarithmic corrections.
These results for $\tau$ are in agreement with the results obtained by Swift 
et. al.\cite{maritan} by using a more indirect mapping to the
age distribution of particles in a related reaction-diffusion process and 
also assuming scaling behavior.  

With $g_0(s)$ completely determined, we therefore have an exact expression
in Eq. (\ref{twentythree}) for $F(\{k\},s)$, the joint Laplace-Fourier
transform of the full correlation function $C(\{x\},t)$ in arbitrary
dimensions. For arbitrary $d$, it is complicated to invert this transform
to obtain an exact expression for $C(\{x\},t)$. However, by analyzing the
small $s$ and small $k$ behavior of $F(\{k\},s)$, it is easy to see that
for large $x$ and large $t$, but keeping $xt^{-1/2}$ fixed, $C(\{x\},t)$
satisfies a scaling behavior, $C(\{x\},t)\sim t^{\gamma}G(\{ {x_i\over
{\sqrt t}} \})$ with $\gamma=2$ for $d=1$ and $\gamma=(3-{d\over 2})$ for
$d>2$ and the scaling function $G(y)$ depends explicitly on $d$.

For $d=2$, there is additional logarithmic correction and the scaling breaks 
down. In this case, the exact expression for $I(s)$ is given by,
\be
I=\frac{2}{\pi(s+4 p)} K(4p/(s+4 p),
\label{twentysix}
\ee
where $K$ is a complete elliptic integral\cite{gradshteyn}.
This gives an explicit expression for $g_0(s)$,
\be
g_0(s)=\frac{\pi (s+4 p)}{s^4 K(4p/(s+4 p))}+\frac{1}{s^2}-\frac{2}{s^3}.
\label{twentyseven}
\ee
Substituting
the expression for $g_0(s)$ in Eq. (\ref{twentythree}) we get,
\be
F(k_1,k_2,s)=\frac{2}{s^3} -\frac{\pi (s+4 p)}{s^3 K(4p/(s+4 p))\left[s+2
p(2-\cos(k_1)-\cos (k_2)) \right]}.
\label{twentyeight}
\ee
After some straightforward algebra, it turns out that the large distance 
behavior of ${\tilde C}(x,y,s)=\int_0^{\infty}C(x,y,t)e^{-st}dt$ is given by,
\be
{\tilde C}(x,y,s)=\frac{-(s+4 p)}{2 s^3 p K(4 p/(s+4 p))}
K_0\left(\frac{r\sqrt{s}}{p},
\right)
\label{thirtyone}
\ee
where $r=\sqrt{x^2+y^2}$ and $K_0$ is the modified Bessel
function\cite{gradshteyn}. In order to get an explicit expression
of $C(x,y,t)$ for large $r$ and $t$, one needs to invert the
Laplace transform given by Eq. (\ref{thirtyone}). But it is obvious
from this expression that $C(x,y,t)$ will no longer have a nice scaling
form in the large distance and long time limit as in one dimension
(see Eq. (\ref{ten})) due to the appearance of logarithms in
the asymptotic behavior of the functions $K(x)$ and $K_0(x)$.
This violation of scaling due to logarithms is again expected since $2$ is
the upper critical dimension of the TM. 

\subsection{Temporal Correlations}
We can write down the equations for the time evolution of the temporal
correlation function in a manner very similar to that in $d=1$.
We define the connected 
correlation function as $D_{x_1,\ldots,x_d}(t,\tau)=\langle 
m_{0,\ldots,0}(t)m_{x_1,\ldots,x_d}(t+\tau)\rangle-t(t+\tau)$. From the 
evolution equation of the masses, it is easy to show that 
the function $D_x(t,\tau)$ evolves as a function of $\tau$ for fixed $t$ as,
\be
\frac{d}{d\tau}D_{\{x\}}(t,\tau)=p\left(\sum_{j=1}^{d}(D_{x_1.\ldots,x_j-1,
\ldots, x_d}(t,\tau)-d D_{\{x\}}(t,\tau)\right) ,
\label{thirtyfour}
\ee
starting from the initial condition $D_{\{x\}}(t,0)=C_{\{x\}}(t)$ where 
$C_{\{x\}}(t)$
is the equal-time correlation function.
Let $H(\{k\},t,\tau)=\sum_{\{x\}=-\infty}^{+\infty} 
D_{\{x\}}(t,\tau) \exp(\sum_{j=0}^{d}i
k_j x_j)$.
From Eq. (\ref{thirtyfour}) we immediately get
\be
H(\{k\},t,\tau)=H(\{k\},t,0)\exp\left(p\tau \sum_{j=1}^{d}(e^{i
k_j}-1)\right),
\label{thirtyfive}
\ee
where $H(\{k\},t,0)$ is the Fourier transform of the equal-time correlation 
function.

We invert Eq. (\ref{thirtyfive}) to get the temporal correlations in terms
of the equal-time correlation functions,
\bea
D_{\{x\}}(t,\tau)&=&\frac{1}{(2 \pi)^d} \int_0^{2 \pi} d^dk H(\{k\},t,\tau) 
\exp(-i \sum_{j=1}^{d}k_j x_j) ,\nonumber\\
&=& \frac{e^{-p\tau d}}{(2\pi)^d}\int_{0}^{2 \pi}d^dk \exp(-i\sum_{j=1}^
{d}k_j x_j)\sum_{\{x^{\prime}\}=-\infty}
^{\infty} C(\{x^{\prime}\},t) \exp(i\sum_{j=1}^{d}k_j x^{\prime}_j)
\exp(p\tau\sum_{j=1}^{d}e^{i k_j}).
\label{thirtysix}
\eea

As in $d=1$, there is a net drift velocity, $p\tau$, in each forward
direction. Hence the correct autocorrelation function is given by
$D_{\{p\tau\}}(t,\tau)$.  Putting $x_j=p\tau$ in Eq. (\ref{thirtysix}), and
simplifying, we get
\be
D_{\{p\tau\}}(t,\tau)=
\sum_{\{m\}=0}^{\infty}C(p\tau-m_1,\ldots,p\tau-m_d)e^{-dp\tau}
\frac{(p\tau)^{\sum_j m_j}} {\prod_j m_j!}.
\label{thirtyseven}
\ee
It can then be shown that in the scaling limit, $\tau\to \infty$, $t\to
\infty$ but keeping $\tau/t$ fixed, the normalized autocorrelation
function allows for a scaling solution as in $d=1$,
\be
A_{\{ p\tau\}}(t,\tau)\sim {1\over {\tau^{d/2}} }h({\tau \over t}).
\label{thirtyeight}
\ee
We do not present the explicit form of the scaling function here. It can
be shown that $h(y)\to {\rm const.}$ as $y\to 0$. Thus in arbitrary
dimensions, the asymptotic decay of the normalized auto-correlation
function is given by, ${\tau}^{-d/2}$ for large $\tau$ (after taking the
large $t$ limit).

\section{Correlations in the \tq-model of Force Fluctuations}
The $q$-model of force fluctuations has been defined in section (II.B). In
this model, the variables $m_i$ evolve in time via the stochastic
equation, $m_i(t+1)=(1-r_i)m_i(t)+r_{i-1}m_{i-1}(t)+1$,
where the random variables $r_i$'s are drawn independently from a 
arbitrary distribution $f(r)$ over the support $[0,1]$.  Experimental results
for the force distribution in real bead packs were found to be described 
accurately by the $q$-model with uniform distribution, $f(r)=1$\cite{liu}.

In this section, we calculate exactly the two point correlations between
$m_i$'s for the generalized $q$-model, i.e., for any arbitrary
distribution $f(r)$. It turns out that the two point correlations are
characterized by two parameters, $\mu_1=\int_0^1 rf(r)dr$ and
$\mu_2=\int_0^1 r^2f(r)dr$ with $\mu_1\geq \mu_2$. In the $\mu_1\geq
\mu_2$ plane, there are two types of asymptotic behaviors depending on
whether $\mu_1=\mu_2$ or $\mu_1> \mu_2$. At every point on the line
$\mu_1=\mu_2$, the correlation function has the same universal asymptotic
behaviour independent of initial conditions as long as they are short
ranged. The special case of the TM with $\mu_1=\mu_2=1/2$ falls within
this class. On the other hand, for all points in the plane $\mu_1>\mu_2$,
the correlation function has once again the same universal asymptotic
behavior regardless of the actual values of $\mu_1$ and $\mu_2$ but this
behavior is different from the behavior on the line $\mu_1=\mu_2$. The
$q$-model with uniform distribution corresponds to the point $\mu_1=1/2$,
$\mu_2=1/3$ and therefore falls in the second category.

Since the steps of the calculation follow closely that of the TM, we will
skip most of the details and present only the final results.

\subsection{Equal-Time Correlations in One Dimension}

Starting from the stochastic evolution equation (Eq.(\ref{two})) of masses in
$d=1$, it is straightforward to write down the evolution equation for the
equal time correlation function, $C_x(t)=\langle m_0(t)m_x(t)\rangle
-t^2$. We find that for general distribution $f(r)$, the equal-time
correlations evolve as
\bea
C_x(t+1)&=&(\mu_1-{\mu_1}^2)( C_{x+1}+C_{x-1})+ (1-2\mu_1+2\mu_1^2)C_x 
\nonumber \\
& & + (\mu_2-{\mu_1}^2) (C_0+t^2) (2\delta_{x,0}-\delta_{x,1}-
\delta_{x,-1}),
\label{corr}
\eea                                     
where $\mu_1=\int_0^1 rf(r)dr$ and $\mu_2=\int_0^1 r^2f(r)dr$. Note that
for the TM, $\mu_1=1/2$ and $\mu_2=1/2$ and then Eq.(\ref{corr}) reduces
exactly to Eq. (\ref{three}) studied in section (III.A). For the uniform
distribution, $f(r)=1$, one gets, $\mu_1=1/2$ and $\mu_2=1/3$.

We can solve the Eq.(\ref{corr}) for arbitrary initial condition and
arbitrary parameters $\mu_1$ and $\mu_2$. It turns out that the dependence
on the initial condition drops out for asymptotically large $t$. Following
the same steps as in the TM in section-III.A, we find that the Laplace
transforms, ${\tilde C}_x(u)=\sum_{t=0}^{\infty} C_x(t)u^t$, are given
exactly by
\bea
{\tilde C}_0(u)&=&\frac{u (1+u)
[\sqrt{1+(4a-1)u}-\sqrt{1-u}]}{(1-u)^3 g_1(u)}, \label{forty}\\
{\tilde C}_x(u)&=&\frac{-u (1+u)}{(1-u)^{5\over 2} g_1(u)} 
\left(\frac{\sqrt{1+(4a-1)u}-\sqrt{1-u}}{\sqrt{4au}}\right)^{2 x},
\label{fortyone}
\eea
where $g_1(u)=[{{a-b}\over b}\sqrt{1+(4a-1)u}+\sqrt{1-u}]$ and
$a=\mu_1-{\mu_1}^2$ and $b=\mu_2-{\mu_1}^2$. It is clear from the above
expressions that the asymptotic behaviors for large $t$ (corresponding to
$u\to 1$) depend on whether $a=b$, i.e., $\mu_1=\mu_2$ or $a>b$, i.e.,
$\mu_1>\mu_2$.

For $\mu_1=\mu_2$, we find, after inverting the Laplace transforms, that for
large $t$, 
\bea
C_0(t)&\approx& { {32{\sqrt a}}\over {15\sqrt{\pi}} }t^{5/2}, \qquad t>>1,
\label{fortytwo} \\
C_x(t)&\approx&-t^2 G\left(\frac{x}{\sqrt{4at}}\right), \qquad
x,t>>1,
\label{fortythree}
\eea
where the universal scaling function $G(y)$ is given by Eq. (\ref{twelve})
as calculated for the TM.

For $\mu_1>\mu_2$, on the other hand, we find for large $t$,
\bea
C_0(t)&\approx& {b\over {a-b} }t^{2}, \qquad t>>1,\label{qas1} \\
C_x(t)&\approx&-{b\over { {\sqrt a}(a-b)}}t^{3\over 2} G_1\left(\frac{x} 
{\sqrt{4at}}\right), \qquad
x,t>>1,
\label{qas2}
\eea               
where the universal scaling function $G_1(y)$ is given by
\be
G_1(y)=\frac{2}{3} \left( \frac{2 e^{-y^2} (1+y^2)}{\sqrt{\pi}}-
y (3+2 y^2) {\rm {erfc}}(y)\right).
\label{fortyfour}
\ee
Note that the uniform distribution corresponds to $\mu_1=1/2$ and
$\mu_2=1/3$,  i.e., $a=1/4$ and $b=1/12$.
In Fig. 2, we compare the numerically obtained scaling plots for the  
uniform distribution with the exact scaling
function given by Eq. (\ref{fortyfour}).

If we define the scaled weight $v=m/t$ as in
Ref.\cite{coppersmith}, then the connected part of the two-point
correlations $\langle v(0,t)v(x,t)\rangle -1 \simeq -t^{-1/2}G_1(x/{\sqrt
{4at} })$ for large $x$ and $t$. Clearly as $t\to \infty$, the scaled weights
$v$'s get completely uncorrelated but for finite depth (or time) $t$,
there is a nonzero anti-correlation specified by the scaling function
$G_1(u)$ which might be possible to measure experimentally. We also point out
that the statement of Claudin et. al.\cite{Claudin} that for the uniform case, 
the correlation is zero at any altitude $t$ is clearly incorrect.

We note that from Eq. (\ref{corr}), one can easily derive the evolution 
equation for the correlation function in the continuum space and time. 
For a proper continuum limit in time, we need to assume that both $a\to 
a\Delta t$ and $b\to b\Delta t$ are of order $\Delta t$. 
Defining the Fourier transform, $G(k,t)=\int_{-\infty}^{\infty}
C(x,t)e^{ikx}dx$, one finds from Eq. (\ref{corr}) that for small $k$, the
correlation function evolves via the equation,
\be
(\partial_t + ak^2)G(k,t)= bk^2\left( \int {dk'\over {2\pi}}G(k',t) 
+t^2\right).  
\label{conti}
\ee
Note that this equation above is identical to the one derived by Claudin 
et. al.\cite{Claudin} except for the additional $k^2t^2$ term on the right 
hand side of Eq. (\ref{conti}). The origin of this additional term can be
traced back to the fact that in our case, the bead mass is nonzero (equal 
to $1$) as opposed to the zero mass case considered in reference 
\cite{Claudin}. As seen from our analysis that for nonzero
mass, the correlation function has a much more nontrivial and universal 
structure as opposed to the rather structure less correlations found in the 
zero mass case in \cite{Claudin}. 

Rescaling time by the factor $a$, one sees that Eq.(\ref{conti}) is
parameterized by the single variable $\lambda=b/a$. As in the discrete
case, there are two possible asymptotic behaviors depending on the value
of $\lambda$. For $\lambda=1$, one finds the Takayasu type behavior and a
completely different asymptotic behavior emerges for all $\lambda<1$.

\subsection{Temporal Correlations in One Dimension}
The two time correlations can similarly be computed for the
$q$-model for any arbitrary distribution $f(r)$.
We present here only the explicit results for the uniform distribution
in $d=1$.

For two time correlations we use the same notation as in the TM (see
section (III.B)). For the uniform distribution, $f(r)=1$, we find that in
the scaling limit, $\tau\to \infty$ and $t\to \infty$ but keeping $\tau/t$
fixed, the normalized auto correlation function, as defined in Eq.
(\ref{norm1}), has the scaling behavior,
\be
A_{\tau/2}(t,\tau)= {1\over {\sqrt \tau}}h_2\left ( {\tau\over {t}}\right ),
\label{scale5}
\ee
where the scaling function is given by,
\be
h_2(y)={\sqrt {2\over \pi}} {1\over {3 (1+y)}}\left [ 3-2\sqrt y (y+2)^{3/2}
+6 y+2 y^2\right ].
\label{scale6}
\ee
The scaling function $h_2(y)\to {\sqrt {2\over \pi}}$ as $y\to 0$ as in
the TM. For large $y$ also, $h_2(y)$ decays as a power law, $h_2(y)\approx
{\sqrt {2\over {9\pi}}}y^{-2}$ as in the TM but with a different exponent
$2$ than for the TM exponent $9/4$.

\subsection{Results for Arbitrary Dimensions}

Following the similar line of arguments as in the TM, the evolution
equation for the two-point correlation function can be derived for the
generalized $q$-model in arbitrary spatial dimension with discrete space
and time. The discrete equations are rather complicated but the asymptotic
properties can be easily derived by taking continuum space and time limit.
In the continuum limit, it turns out that as in $d=1$, the Fourier
transform in any dimension, $G(\{k_i\}, t)=\int_{-\infty}^{\infty}
C(\{x_i\},t)\exp(i\sum k_ix_i) d^dx$ evolves by the simple equation
(\ref{conti}) parameterized by the ratio $\lambda=b/a$ once time is
rescaled by $a$. The parameter $\lambda\leq 1$ as in the discrete case. As
in the TM, taking the Laplace transform of Eq. (\ref{conti}) with respect
to $t$, we find
\be
F(\{k\},s)={ {\lambda k^2[2+g_0(s)s^3]}\over {s^3(k^2+s)}},
\label{ad1}
\ee
where $F(\{k\},s)=\int_0^{\infty}G(\{k\},t)e^{-s t}dt$ and
$g_0(s)=\int_0^{\infty}C(0,\ldots,0,t)e^{-s t}dt$ which is yet to be
determined. As in the TM, $g_0(s)$ is determined by integrating Eq.
(\ref{ad1}) with respect to $k$. Note that the upper cut-off of each $k_i$
integration is now set by $2\pi/{\Lambda}$ where $\Lambda$ is the lattice
constant. We find
\be
g_0(s)= { {2\lambda[1-sI(s)]}\over {s^3[1-\lambda +\lambda sI(s)]}},
\label{ad2}
\ee
where $I(s)= \int d^dk/(k^2+s)$. The small $s$ behaviour of $I(s)$ is same
as given by Eq. (\ref{Is}).

From Eq. (\ref{ad2}) it is clear that there are two different small $s$
behaviors of $g_0(s)$ depending on whether $\lambda=1$ or $\lambda<1$. The
TM corresponds to $\lambda=1$ and its asymptotic behaviors have already
been discussed in detail in Section-(IV). Here we focus on $\lambda<1$. In
that case, using the small $s$ behavior of $I(s)$ in Eq. (\ref{ad2}), it
is clear that $g_0(s)\sim 2\lambda/s^3$ in any dimension, indicating that
$C(0,0,\ldots,0,t)\sim t^2$ for large $t$. Thus for $\lambda<1$, in contrary 
to the $\lambda=1$ case (TM), there is no critical dimension separating 
different asymptotic growth of $C(0,0,,\ldots,0,t)$.

Substituting this expression for $g_0(s)$ in Eq. (\ref{ad1}), we find that
for small $s$
\be
F(\{k\},s)\approx { {2\lambda k^2}\over { (1-\lambda)s^3(k^2+s)}}.
\label{ad3}
\ee
It is then not difficult to derive the asymptotic properties of the
correlation function in real space and time. We find,
\bea
C(\{0\},t)&\sim& t^2, \qquad t>>1,\label{ad4} \\
C(\{x\}, t)&\approx&-t^{(2-{d\over 2})} G_1\left(\frac{x}{\sqrt{t}}\right),
\qquad x,t>>1,
\label{ad5}
\eea   
where $G_1(y)$ is the dimension dependent scaling function. 

The most important result of this subsection is that while for
$\lambda=1$, there is an upper critical dimension $d_c=2$ separating
different asymptotic growth of the on site variance, there is no such
critical dimension for $\lambda<1$ (which includes the uniform distribution 
of fractions for the $q$-model).

\section{Summary and Conclusion}
In this paper we have computed exactly both the equal-time as well as the
unequal-time two point correlations in the Takayasu model of mass
aggregation and injection in all dimensions. We have identified different
scaling limits and obtained the scaling functions explicitly in $d=1$. Our
exact results for the on-site mass variance prove rigorously, without any
assumption of scaling, that the upper critical dimension of the Takayasu
model is $d_c=2$.

We have also extended our technique to compute exactly the correlations in
a larger class of aggregation models with injection. This generalized
model includes, as special cases, the Takayasu model and also the
$q$-model of force fluctuations in granular materials. We have shown that
the correlation functions in this generalized model is parameterized by two
variables $(\mu_1,\mu_2)$ which are respectively the first and the second
moment of the distribution $f(r)$ of the fractions $r_i$'s. We have shown
that in the two-dimensional parameter space $(\mu_1,\mu_2)$ with
$\mu_1\geq \mu_2$, there are two types of asymptotic behaviors of the
correlation function depending on whether $\mu_1=\mu_2$ or $\mu_1 >
\mu_2$.  For generic points in the region $\mu_1>\mu_2$ which includes the
uniform distribution represented by $(1/2,1/3)$, the correlations have
similar asymptotic behaviors which is different from that on the line
$\mu_1=\mu_2$ which includes the Takayasu model. Besides, for
$\mu_1>\mu_2$, there is no upper critical dimension in contrast to the
case $\mu_1=\mu_2$ where the upper critical dimension is $d_c=2$. We have
presented explicit forms of scaling functions for both the Takayasu line
as well as the experimentally relevant uniform distribution case.  These
exact results will be useful for comparison with possible future
experimental results on correlations in bead packs.

In this paper we have calculated exactly various time-dependent
correlations between forces in bead packs in the context of the simple
scalar $q$-model. There have been various generalizations of this scalar
model to include the tensorial nature of the forces\cite{Claudin,copper2}
and also to non-cohesive granular materials\cite{socolar}. It would be
interesting to see if our method can be extended to calculate the
correlations in these generalized models.

We thank D. Dhar and M. Barma for useful discussions.

\newpage
\begin{figure}
\begin{center}
\leavevmode
\psfig{figure=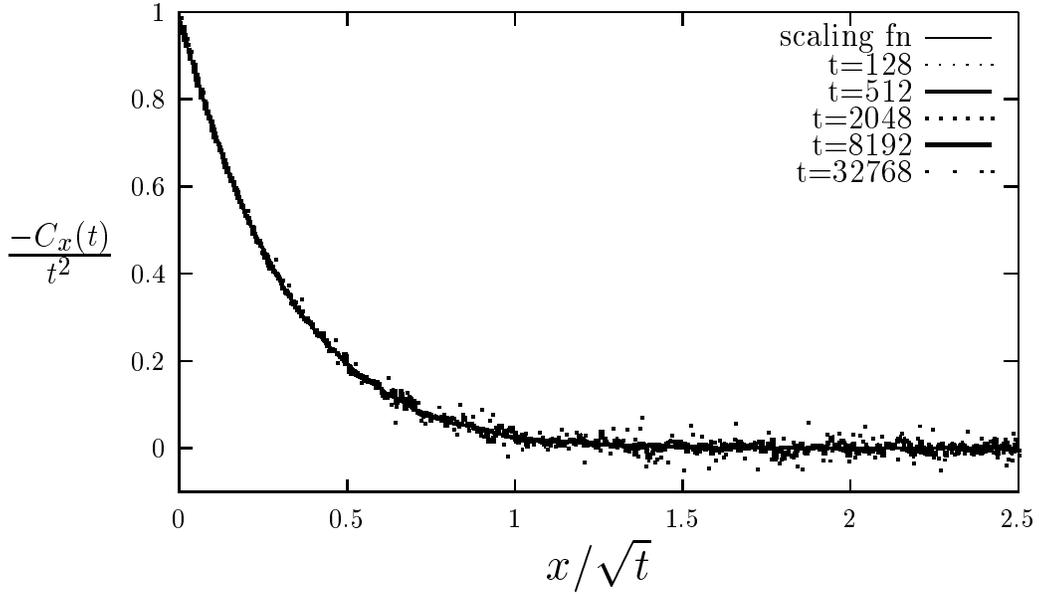,width=14cm,angle=0}
\caption{The figure shows the scaling plots of
the equal-time correlation function, $C_x(t)=\langle m(0,t)m(x,t)\rangle
-t^2$ obtained from numerical simulation of the TM on a one dimensional
lattice of $1000$ sites. The data for five different times collapse
onto a single scaling curve which matches very well with the
analytical scaling function given by Eq. (\ref{twelve}).}
\end{center}
\end{figure}

\begin{figure}
\begin{center}
\leavevmode
\psfig{figure=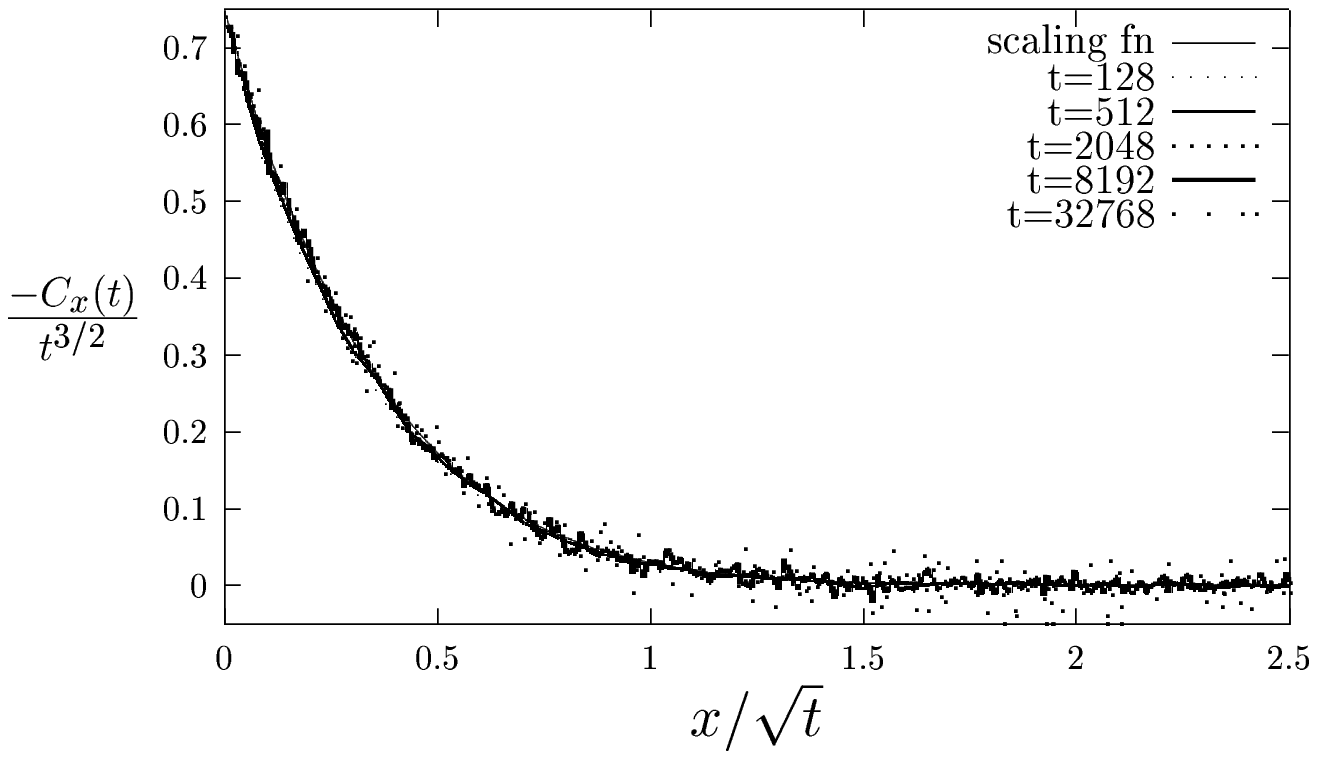,width=14cm,angle=0}
\caption{The figure shows the scaling plots of
the equal-time correlation function, $C_x(t)=\langle m(0,t)m(x,t)\rangle
-t^2$ obtained from numerical simulation of the $q$-model on a one dimensional
lattice of $1000$ sites. The data for five different times collapse
onto a single scaling curve which matches very well with the
analytical scaling function given by Eq. (\ref{fortyfour}).}
\end{center}
\end{figure}

\end{document}